\documentclass[twoside,twocolumn,9pt]{article}
\usepackage{extsizes}
\usepackage[super,sort&compress,comma]{natbib} 
\usepackage[version=3]{mhchem}
\usepackage[left=1.5cm, right=1.5cm, top=1.785cm, bottom=2.0cm]{geometry}
\usepackage{balance}
\usepackage{mathptmx}
\usepackage{amssymb}
\usepackage{sectsty}
\usepackage{graphicx} 
\usepackage{lastpage}
\usepackage[format=plain,justification=justified,singlelinecheck=false,font={stretch=1.125,small,sf},labelfont=bf,labelsep=space]{caption}
\usepackage{float}
\usepackage{fancyhdr}
\usepackage{fnpos}
\usepackage[english]{babel}
\addto{\captionsenglish}{%
  
}
\usepackage{array}
\usepackage{droidsans}
\usepackage{charter}
\usepackage[T1]{fontenc}
\usepackage[usenames,dvipsnames]{xcolor}
\usepackage{setspace}
\usepackage[compact]{titlesec}
\usepackage{hyperref}

\usepackage{epstopdf}

\definecolor{cream}{RGB}{222,217,201}

\begin{document}

\pagestyle{fancy}
\thispagestyle{plain}
\fancypagestyle{plain}{
\renewcommand{\headrulewidth}{0pt}
}

\makeFNbottom
\makeatletter
\renewcommand\LARGE{\@setfontsize\LARGE{15pt}{17}}
\renewcommand\Large{\@setfontsize\Large{12pt}{14}}
\renewcommand\large{\@setfontsize\large{10pt}{12}}
\renewcommand\footnotesize{\@setfontsize\footnotesize{7pt}{10}}
\makeatother

\renewcommand{\thefootnote}{\fnsymbol{footnote}}
\renewcommand\footnoterule{\vspace*{1pt}%
\color{cream}\hrule width 3.5in height 0.4pt \color{black}\vspace*{5pt}} 
\setcounter{secnumdepth}{5}

\makeatletter 
\renewcommand\@biblabel[1]{#1}            
\renewcommand\@makefntext[1]%
{\noindent\makebox[0pt][r]{\@thefnmark\,}#1}
\makeatother 
\renewcommand{\figurename}{\small{Fig.}~}
\sectionfont{\sffamily\Large}
\subsectionfont{\normalsize}
\subsubsectionfont{\bf}
\setstretch{1.125} 
\setlength{\skip\footins}{0.8cm}
\setlength{\footnotesep}{0.25cm}
\setlength{\jot}{10pt}
\titlespacing*{\section}{0pt}{4pt}{4pt}
\titlespacing*{\subsection}{0pt}{15pt}{1pt}

\renewcommand{\topfraction}{0.9}       
\renewcommand{\bottomfraction}{0.9}    
\renewcommand{\textfraction}{0.05}     
\renewcommand{\floatpagefraction}{0.9} 


\fancyhead{}
\renewcommand{\headrulewidth}{0pt} 
\renewcommand{\footrulewidth}{0pt}
\setlength{\arrayrulewidth}{1pt}
\setlength{\columnsep}{6.5mm}
\setlength\bibsep{1pt}

\makeatletter 
\newlength{\figrulesep} 
\setlength{\figrulesep}{0.5\textfloatsep} 

\newcommand{\topfigrule}{\vspace*{-1pt}%
\noindent{\color{cream}\rule[-\figrulesep]{\columnwidth}{1.5pt}} }

\newcommand{\botfigrule}{\vspace*{-2pt}%
\noindent{\color{cream}\rule[\figrulesep]{\columnwidth}{1.5pt}} }

\newcommand{\dblfigrule}{\vspace*{-1pt}%
\noindent{\color{cream}\rule[-\figrulesep]{\textwidth}{1.5pt}} }

\makeatother

\newcommand{\gdot}{\dot{\gamma}}
\newcommand{\invgdot}{\dot{\gamma}^{-1}}
\newcommand{\rate}[2]{\dot{\gamma} = #1 \times 10^{#2} \;\; \tau^{-1}}


\twocolumn[
  \begin{@twocolumnfalse}

\LARGE{\textbf{Disentangling microstructural elements of shear thickening suspensions via computer simulations of a minimal model}} \\
\vspace{0.3cm}

\noindent\large{William C. J. Buchholtz,\textit{$^{a,b}$}, Daniel L. Blair,\textit{$^{a,b}$} Jeffrey S. Urbach,\textit{$^{a,b}$}, H. A. Vinutha\textit{$^{a,b}$} and Emanuela Del Gado\textit{$^{a,b}$}} \\

\begin{center}
\begin{minipage}{0.85\textwidth}
\normalsize{We use a minimal model for a dense suspension undergoing thickening and thinning to investigate microstructural changes in $2d$ simulations. Our simulations show that in steady flow the contact network contains distinct building blocks which are clearly signaled by sharp peaks in the radial distribution function, similar to what is observed in granular jamming. These structures {deform} during thinning. Non-Gaussian stress fluctuations that only emerge during thickening are associated to power law tails in the distribution of local contact forces, which tend to emerge when the flow-induced building blocks form large spanning assemblies. The subset of the contact network characterized by strong contact forces and connectivity large enough to be rigid or over-constrained is increasingly likely to percolate as the system starts to thicken, and to percolate over larger strain windows during thickening. The tendency of these structures to span the sample and to persist is dramatically reduced during thinning, where instead their {deformation} allows for a more homogeneous spatial redistribution of contact forces, significantly reducing the fluctuations of the macroscopic stress over time.} 
\end{minipage}
\end{center}




 \end{@twocolumnfalse} \vspace{0.6cm}

]

\renewcommand*\rmdefault{bch}\normalfont\upshape
\rmfamily
\section*{}
\vspace{-1cm}

\footnotetext{\textit{$^{a}$~Department of Physics, Georgetown University, 3700 O St, NW  Washington,  DC 20057, USA.}}
\footnotetext{\textit{$^{b}$~Institute for Soft Matter Synthesis and Metrology, Georgetown University, 3700 O St, NW  Washington,  DC 20057, USA.}}




 \section{Introduction}

The proliferation of frictional and complex surface contacts between particles has been recognized as an essential ingredient in the build up of shear stress in dense suspensions under shear \cite{Seto_2013,Mari_2014,Wyart_2014,Cohen_2015,Singh_2018,Morris_Review,Del_Gado_2020,Moghimi_2025,Moghimi_202597,Moghimi_2024omn,Miller_2022}. Shear thickening of particle suspensions is also often accompanied by large-scale spatial inhomogeneities of the shear stress, under an imposed rate, which also features anomalous fluctuations in time, detected in experiments and simulations \cite{Boersma_1991,Lootens_2003,Lootens_2004,Heussinger_2013,Mari_2014,Rathee_2017,Ong_2020,Rathee_2020,Rathee_2022,Gauthier_2023,Goyal_2024}. Large-scale spatio-temporal heterogeneity of the stress-bearing part of the contact network has been hypothesized to explain these observations. Further insight into the growth and the morphology of the frictional contact network during shear thickening, therefore, can shed new light on the nature of the shear thickening transition and on how stresses can be controlled during flow. 

In dry granular assemblies undergoing jamming or shear jamming, several studies have highlighted and clarified the distinct roles that force and contact networks play, as minimal changes in the contact network correspond to large-scale reorganization of the force network \cite{Nott2020Coherent}, and over-constrained portions of the contact network have been shown to percolate in correspondence with the shear jamming transition \cite{Cates_1998,Bi_2011,Papadopoulos_2018,Vinutha_2019,Vinutha_Nature,Deng_2023}. Both the jamming and the shear jamming transitions have been connected to the scaling properties of shear thickening transitions in dense suspensions \cite{Wyart_2014,Ramaswamy_2023}, therefore suggesting that similar roles of force and contact networks may emerge also there. As a consequence, there has been significant interest in characterizing how the network of frictional contacts changes in size and topology during thickening \cite{Gameiro_2020,Sedes_2022,Nabizadeh2022structure,van_der_Naald_2023,DAmico_2025} and, in particular, which subset of the frictional contact network is most responsible for the changes in the suspension rheology when it thickens \cite{Goyal_2024,DAmico_2025}. 

To address the many open questions on the growth and morphology of contact networks in shear thickening suspensions, computational studies in 2d and 3d have analyzed subsets of the frictional contact network and their evolution during thickening. One interesting subset has been identified through groups of particles that all have at least $k$ frictional contacts with members of the same group ($k$-cores). In 3d, the presence of cores with $k \leq 3$ was found to display significant correlation with the increase in the average stress with increasing the shear rate \cite{Sedes_2022, DAmico_2025}. Locally rigid clusters of particles with at least 4 frictional contacts in 3d (4-neighbor clusters) have been shown to undergo a percolation transition during thickening, following a finite-size-scaling ansatz similar to equilibrium critical phenomena \cite{Goyal_2024}. As $4$ contacts in 3d provides enough constraints, in presence of friction, to guarantee isostaticity within a mean-field Maxwell criterion \cite{Henkes_2016}, these clusters are expected to be locally rigid or overconstrained when the contacts are not lubricated. We note that the $4$-neighbor clusters are different from the $4$-cores, for example, in that not all of the particle contacts need to have at least $4$ contacts as well, so they tend to identify the backbone of $4$-cores. Close to the percolation transition of these clusters, their percolation/de-percolation during steady flow seems to coincide with giant fluctuations of the stress \cite{Goyal_2024}, potentially providing a microscopic explanation of several aspects of the shear thickening phenomenology reported in experiments \cite{Rathee_2017,Ong_2020,Miller_2022,Ramaswamy_2023,Moghimi_2025}. 

In other recent studies, two-dimensional simulations have shown that fully shear thickened flow states at high stress  always feature system spanning clusters that are rigid according to the pebble game algorithm, which identifies clusters of particles with enough constraints to satisfy the isostaticity condition \cite{van_der_Naald_2023}. It has also been shown that proliferation and growth of such clusters is always favored when large enough stresses are imposed, and  close to jamming features several characteristics typical of critical phenomena \cite{Santra_2024, Pandare2026contact}. 
Collectively, all these studies support the idea that flow at high rates induces some level of self-organization of particle contacts which then, in turn, changes stress transmission and rheological responses. More investigation is however needed to fully characterize specific microstructures that control stress transmission and their statistical properties. Recent studies of activity-driven shear thickening in model active suspensions have also highlighted how system spanning networks of constraints may appear even in model active suspensions, akin to strongly far-from equilibrium, self-propelled living systems \cite{Bayram_2023}. In sum, gaining further insight into the self-organization underpinning shear thickening would not only open the path to truly control and design the flow of a  wide range of particulate dense suspensions, but would also bring new knowledge and understanding of non-equilibrium, strongly driven, many body physical systems in broad sense. 

One of the first outstanding questions is how certain microscopic structures developed under flow are specific of shear thickening, as self-organization under flow in dense suspensions is a very broad phenomenon that does not only occur in those conditions. However, separating and disentangling microstructural motifs that emerge in different conditions is very difficult even in numerical simulations that have full access to microstructures, because of the microstructure complexity, especially in three dimensions. 

In this study, we resort to a two-dimensional minimal model for dense suspensions, previously introduced by Grob, Heussinger, and Zippelius \cite{Heussinger_2013} and also studied by Maiti and Heussinger \cite{Maiti_2016}, to specifically identify how subsets of the frictional contact networks formed under shear flow may be substantially different during thickening and thinning, and how they may be related to giant stress fluctuations akin to those detected in experiments. The simplicity of the model chosen, and of the analysis of two-dimensional simulations, allow us to draw interesting general conclusions that may apply to a broad range of systems. First, our simulations show that in steady flow the contact networks of a thickening suspension contain distinct building blocks consisting of small groups of disks. These are clearly signaled by sharp peaks in the radial distribution function which has similarities to what is observed in granular jamming. These structures cannot remain stable and {deform} during thinning. Second, we show that giant, non-Gaussian stress fluctuations that only emerge during thickening are associated to power law tails in the distribution of local contact forces, which tend to emerge when the flow-induced building blocks form large spanning assemblies. The subset of the contact network characterized by strong contact forces (i.e., in the tail of the distribution) and connectivity large enough to be locally rigid or over-constrained (i.e., at least 3 contacts in 2d with frictional interactions) is increasingly likely to percolate as the system starts to thicken, and to percolate over larger strain windows during thickening. We call these structures 3-SFN. Their tendency to span the sample and to persist is dramatically reduced during thinning, where instead their {deformation} allows for a more homogeneous spatial redistribution of contact forces, significantly reducing the fluctuations of the macroscopic stress over time. 

{ 
The fact that the model features both thickening and thinning regimes allows us to analyze microstructures that correspond to the same (or very similar) relative viscosity, for the same solid fraction and contact laws, while being respectively in the thickening vs thinning regime. As a consequence, our work highlights that the relative viscosity (i.e. the average stress) in itself does not contain all the information about how the microstructure affects the rheology of a suspension. More specifically, we obtain new mechanistic insight into how the distribution of angles between contact bond vectors is distinctly different between the thickening and the thinning regimes. By performing simulations with a system size significantly larger than what utilized in previous literature \cite{Mari_2014,Heussinger_2013,Singh_2018,Nabizadeh2022structure}, we show that stress fluctuations developed under equivalent conditions (specific contact laws, particle volume fraction, and shear rate) are qualitatively different in small samples, which display shear stresses oscillating between two values \cite{Heussinger_2013, Singh_2018}, whereas large enough system sizes allowed us to sample a broader distribution throughout a comparable strain window, which is closer to experimental observations \cite{Ong_2020,Rathee_2017,Rathee_2022}. Small system sizes also do not allow for the statistics required to resolve the correlation of the percolation of the 3-SFN clusters with the rheology, which instead can be captured in sufficiently large samples. The 3-SFN, moreover, are a new and complementary diagnostic for characterizing the microstructure, which points to the microscopic origin of the stress fluctuations detected in experiments \cite{Ong_2020,Rathee_2017,Rathee_2020}, akin to the $k$-neighbor clusters introduced in the $3d$ numerical studies \cite{Goyal_2024}. While supporting the picture of a rigidity percolation of $k$-neighbor clusters, we also discuss the connection between 3-SFN and the rigid clusters identified by the pebble game algorithm in $2d$ simulations \cite{van_der_Naald_2023,Santra_2024}.}

The paper is organized as follows: in Section \ref{Methods} we describe the model and the numerical simulations, and Sections \ref{rheology}-\ref{strong-force} discuss the rheological properties measured and the analysis of the structure and contact networks, whereas Section \ref{conclu} contains some concluding remarks and an outlook.

\section{Methods}
\label{Methods}

We consider a two-dimensional (2D) assembly of non-Brownian soft disks sheared at a constant shear rate $\gdot$. The disks are immersed in an implicit solvent of viscosity $\eta_{sol}$, from which they experience a drag force proportional to their velocity relative to the imposed flow. To avoid crystallization, we consider a mixture of two sizes, big (B) and small (S) with diameters $d_B$ and $d_S$ and a diameter ratio $d_B / d_S = 1.4$. The packing fraction $\phi$ of the disks is fixed at $0.79$, which is significantly below the isotropic jamming density of $\phi_J \approx 0.845$ \cite{OHern_2003}.

The disks interact via contact forces described by the Cundall-Strack model \cite{Cundall_1979} as shown in Equation \ref{contact_forces}, where contact forces between disks $i$ and $j$ are calculated using a combination of viscous and elastic terms along both the normal and tangential directions.

\begin{equation}
\label{contact_forces}
\vec{F}_c^{ij} = (\kappa_n \delta - m_{eff} \zeta_n v^{n}_{ij}) \hat{n}_{ij} - (k_t \Delta s + m_{eff} \zeta_t v^{t}_{ij}) \hat{t}_{ij}
\end{equation}

Here the unit vector $\hat{n}_{ij}$ points from particle $i$ to $j$ and $v^{n}_{ij}$ is the component of the relative velocity along that direction. $v^{t}_{ij}$ is found by calculating the relative velocity of the disks $\vec{v}_{ij} = \vec{v}_i - \vec{v}_j$ and taking the component along the direction $\hat{t}_{ij}$ perpendicular to $\hat{n}_{ij}$. $\kappa_n$ and $\kappa_t$ are spring constants acting along the normal (n) and tangential (t) directions respectively while $\zeta_n$ and $\zeta_t$ are the related damping coefficients. The effective mass is given by $m_{eff} = m_i m_j / (m_i + m_j)$. The normal overlap is calculated simply as $\delta = R_{ij} - r_{ij}$ where $R_{ij}$ is the sum of the radii equal to $(d_i + d_j)/2$ and $r_{ij}$ is the inter-particle center of mass separation. The frictional (tangential) force is a function of how far disks have slid past each other since establishing contact. This requires integrating the tangential velocity from the time of contact $\tau_0$:

\begin{equation}
\label{overlap_int}
\Delta s (\tau) = \int_{\tau_0}^\tau v_{ij}^{t} (\tau') d\tau'
\end{equation}

\noindent The maximum value of the tangential force $\vec{F}^{ij}_t$ is given by the Coulomb criterion $F^{ij}_t \leq \mu F^{ij}_n$, where  $F^{ij}_n$ is the normal force, and $\mu$ is the friction coefficient. We choose $\mu = 1.0$ as in other studies \cite{Morris_Review,Maiti_2016}.

The translational and rotational degrees of freedom of the disks are described by the following equations of motion: 

\begin{equation}
\label{translational_motion}
m_i \frac{d \vec{v}_i}{dt} = \sum_j \vec{F}_c^{ij} -\zeta_{sol} \vec{v}_{i,f}
\end{equation}

\begin{equation}
\label{rotational_motion}
I_i \frac{d^2 \theta_i}{dt^2} = \sum_j F_{c,t}^{ij} \frac{d_i}{2} - \zeta'_{sol} \frac{d \theta_i}{dt}
\end{equation}

Here $\vec{v}_{i,f}$ is the velocity of disk $i$ ($\vec{v}_i$) relative to the solvent flow at its location ($\vec{v}_f$), $I_i$, is the moment of inertia about its center of mass, and $\vec{F}_c^{ij}$ is the contact force due to particle $j$ whose tangential component is given by $F_{c,t}^{ij}$. The second term in Eqn. 3 represents a drag force which is proportional to the  difference between the particle velocity $\vec{v}_i$ and the flow velocity $\vec{v}_f(r_i)=\hat{e}_x \dot{\gamma} y_i$. The drag coefficient $\zeta_{sol}$ represents the viscosity of the surrounding fluid, $\zeta_{sol} \sim \eta_{sol}$. The angular velocities are damped by the drag coefficient $\zeta'_{sol}$. 

 We note that the 2D nature of the model, the oversimplified description of the solvent, especially the lack of lubrication forces, and the lack of explicit boundaries in the simulations, limit the possibilities of direct and quantitative comparison with experiments. Nevertheless, as discussed in the following, the study performed did allow us to develop new explanations and to provide new insights potentially relevant to experiments.

In our simulations, lengths are reported as multiples of the small disk diameters $d_S$ while the energy unit $\epsilon$ is equal to the elastic energy of two small disks undergoing full overlap ($\kappa_n d_S^2/2$). 
All disks are given the same mass $m$. Based on these fundamental mass ($m$), energy ($\epsilon$) and length ($d_S$) units, the time unit is calculated as $\tau = \sqrt{m d_S^2/ \epsilon}$, which is approximately equal to the period disks oscillate with when pushed together. The input parameters in our simulations are $\kappa_n = \kappa_t = 2$, $\zeta_n = 3 \sqrt{\epsilon/m}/d_S$, and $\zeta_t = \zeta_n/2$. The solvent damping coefficients ($\zeta_{sol}=0.1 \; m / \tau$ for translations and $\zeta'_{sol} = 0.33 \; (m d_S) / \tau $ for rotations) are chosen such that the inertial forces are not important for the small shear rates considered here. The inertial number $I=\dot{\gamma} d_{S} / \sqrt{P/\rho}$ (where P is the system pressure and $\rho$ is the particle density)  is generally smaller than $O(10^{-3})$.


As discussed by Mari et al. \cite{Mari_2014}, shear thickening inherently requires a time scale in the system that competes with the one corresponding to the inverse shear rate $\invgdot$. For our system, this can be provided by the time scale $\tau_{elastic} = \sqrt{m / \kappa_N}$ of the oscillations that the disks experience along the normal direction after establishing overlap. Comparable time scales can be computed from the tangential spring constant, the two damping constants $\zeta_n$ and $\zeta_t$, and the intrinsic solvent viscosity $\eta_{sol}$. The absence of other  microscopic time scales implies that the softness of the particles must become important in the limit of large shear rate, which leads to a transition from shear thickening to shear thinning as discussed in the following and as also observed in a previous study of the same model \cite{Maiti_2016}. 

From the disks relative positions and forces, we compute the stress tensor

\begin{equation}
\hat{\sigma}=\frac{1}{A}\sum_{i\neq j} \vec{r}_{ij} \otimes \vec{f}_{ij}
\end{equation}

\noindent where $\vec{r}_{ij}$ is the vector from the center of disk $i$ to the point of contact with disk $j$, $\vec{f}_{ij}$ is the contact force, and $A$ is the area of the system \cite{Bi_2011,Azema_2012}. Once the stress tensor is obtained, we measure the shear stress as the off-diagonal component $\hat{\sigma}_{xy}$, referred to subsequently as $\sigma$.

Data is collected for a system size of 20,000 particles. A constant shear rate is applied to each configuration until the system reaches a steady state. For each shear rate we use a statistically independent initial condition. Once steady state is reached, an additional $\Delta \gamma = 10$ is applied during which the system is sampled at even strain intervals of $\Delta \gamma = 0.001$ for a total of 10,000 configurations. All simulations are performed in LAMMPS \cite{Lammps} and with Lees-Edwards periodic boundary conditions. 

\section{Rheology And Fluctuations}
\label{rheology}
\begin{figure}
    \includegraphics[width=\columnwidth]
    {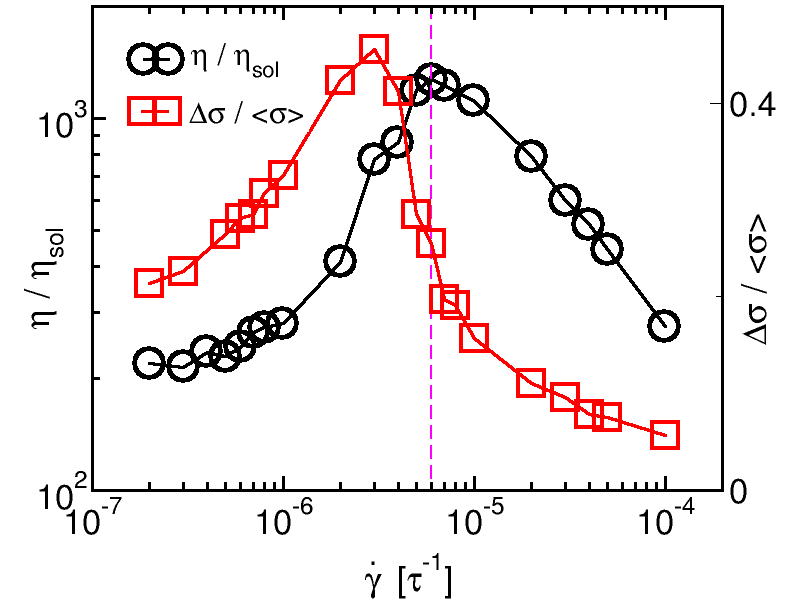}
    \caption{Relative viscosity versus shear rate for a system size of 20,000 particles (black data). The corresponding stress fluctuations in steady state  given as the standard deviation of the stress normalized by the mean (red data). The flow curve exhibits {quasi-Newtonian (QN)} flow at the lowest rates $\dot{\gamma} \lesssim 6 \times 10^{-7} \;\; \tau^{-1}$. Above this regime the system thickens until $\dot{\gamma} \approx 6 \times 10^{-6} \;\; \tau^{-1}$ where thinning is observed due to particle overlaps. The stress fluctuations increase during the thickening but begin to decrease before the start of the subsequent thinning regime.}
    \label{flowcurve}
\end{figure}

Figure \ref{flowcurve} shows the flow curve computed from the simulations once the steady state flow is reached at different shear rates. The range of rates we explore encompasses several orders of magnitude during which the system exhibits {quasi-Newtonian (QN)}, thickening, and thinning flow respectively as the shear rate is increased. Although the shear stress in the system increases monotonically with shear rate, thinning is made possible by the softness of the disks which experience larger overlaps at faster rates. This model choice allows us to directly compare the structures of the same initial microscopic configuration undergoing thickening and thinning. Overall the rheology we report here is both qualitatively and quantitatively in good agreement with the results reported by Maiti et al. \cite{Maiti_2016}, which used the same interaction model for similar densities. The shape of the flow curve does not change significantly by changing the system size, although we note that both the onset of thickening and the onset of thinning seem to move to slightly lower shear rates for smaller samples (see Supplementary Information, Figure 1). 

Together with the relative viscosity, which is obtained from the average shear stress, we plot also the fluctuations of the shear stress, indicating that thickening is characterized by a significant increase in the shear stress fluctuations, whereas the onset of thinning features their significant decrease. The dashed line in the figure identifies the peak of the stress fluctuations while the shadowed region indicates the transition from thickening to {QN} flow (left of the dashed line) or thinning (right of the dashed line). The location of the peak and the width of the thickening and thinning regimes depend on the friction coefficient, the viscosity of the solvent, the Young moduli of the particles, and the shear rate. 
Increasing the stiffness of the
spring used to describe the particle contacts in the model shifts the onset of thinning to higher strain rates \cite{Heussinger_2013}, without changing significantly the overall behavior.
{In the following, shear rates are reported as a fraction of the peak shear rate at which the system begins to thin $\dot{\gamma}' = \dot{\gamma} / \dot{\gamma}_P$  ($\dot{\gamma}_P= 6 \times 10^{-6} \; \tau^{-1}$), for the specific set of parameters used here (see also Supplementary Information for a comparison of two different system sizes).}

\begin{figure}

  \includegraphics[width=\columnwidth]{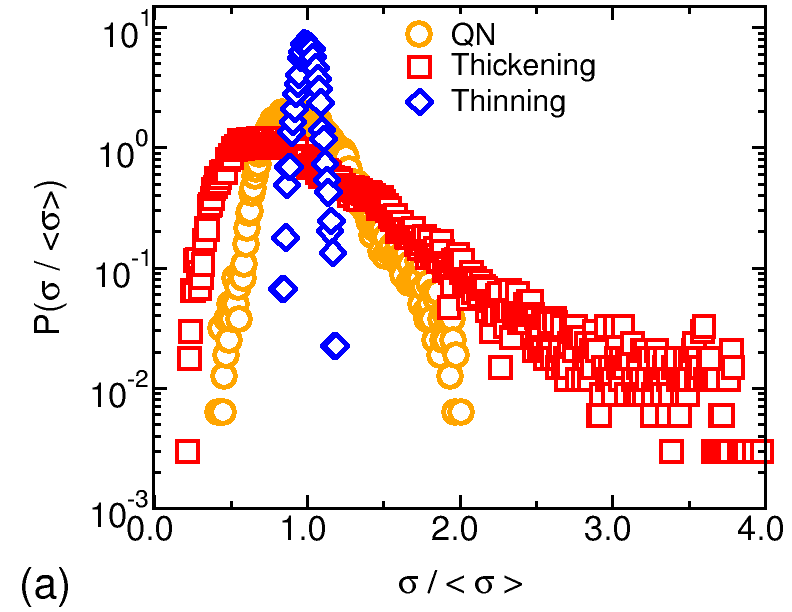}%
  \hfill
  
  \includegraphics[width=\columnwidth]{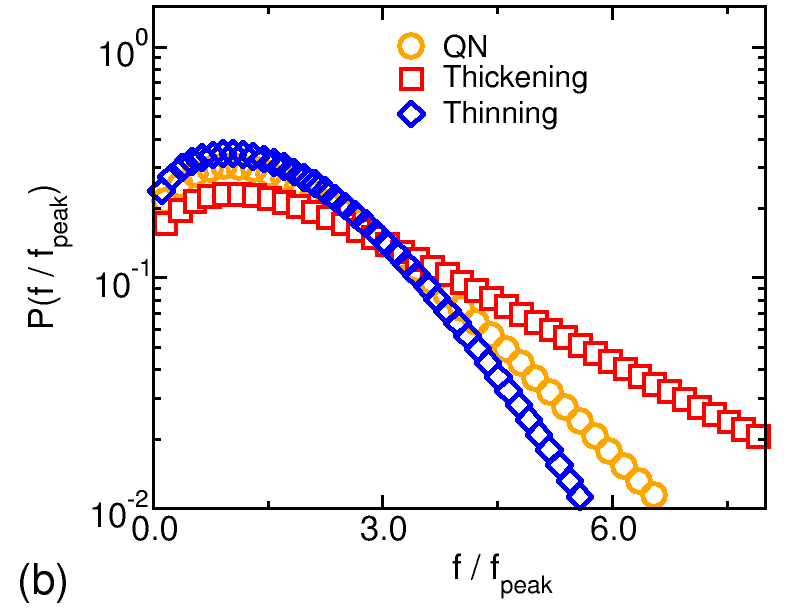}%

\caption{The steady state distributions for the system-wide shear stress $\sigma$ (a) and individual contact forces $f$ (b). {Shear rates are taken from the QN regime ($\dot{\gamma}'=5.0 \times 10^{-2}$), , the thickening ($\dot{\gamma}'=5.0 \times 10^{-1}$), and the thinning ($\dot{\gamma}'=1.7 \times 10^{-1}$).} $\langle \sigma \rangle$ and $f_{peak}$ are the average value of the system shear stress and the peak (most probable) value of the force distribution respectively. Shear rates are chosen from the three observed flow regimes ({QN}, thickening, and thinning). We find that the thickening flow is associated with an increase in fluctuations, in agreement with both  experimental \cite{Boersma_1991, Lootens_2003, Qin_2020} and computational \cite{Heussinger_2013, Mari_2014} studies.}
\label{fluctuations}
\end{figure}

Figure \ref{fluctuations}(a) displays the probability density distributions of the shear stress, obtained from the steady state flow, during the different flow regimes. Each data point indicates the value of $\sigma$ at a given time (or strain) during steady state flow, in the different regimes of the flow curve. During thickening the distribution features a pronounced high stress tail (red squares), and shrinks back dramatically once the system transitions from thickening to thinning. The question of whether and how the large stresses in thickening stem from microscopic changes can be addressed in the simulations: the behavior of the stress fluctuations appears to originate from the fluctuations in the contact forces, as shown by the change in the contact forces distributions going into thickening and thinning flows in Figure \ref{fluctuations}(b). Each data point here corresponds to a specific contact force present across all microscopic configurations sampled during the steady flow, computed over the same time series used in Figure \ref{fluctuations}(a). During thickening, we detect the concomitant growth of fluctuations of microscopic quantities such as the average contact number, which also decrease during thinning (not shown). We note that the comparatively large system size used here allowed us to access the full distributions of shear stress and microscopic contact forces. Other computational studies with smaller system sizes showed instead bimodal distributions \cite{Heussinger_2013, Mari_2014,Seto_2019,van_der_Naald_2023}, which we also recover in small samples (see Supplementary Information, Figures 2 and 3). Overall, these findings offered motivation to investigate the presence of specific structural motifs that could be associated to the changes in the distributions of microscopic contact forces, and in particular to the development of anomalously large contact forces, as explained in the following. 

Notably, we observe that the stress fluctuations (relative to the mean) stop increasing \emph{before} the thickening is over. As well, the thinning regime coincides with a significant reduction in the size of the fluctuations. Frictional contact network percolation is a mechanism which could explain the behavior of stress fluctuations during thickening observed in our model. Networks of frictional disks that percolate across the system are self supporting (due to the periodic boundary conditions) and should be able to carry larger loads than non-percolating networks. If frictional network percolation were a driver of thickening, then thickening could be expected to occur as long as these networks could increase the amount of time they percolated at higher rate. However, \textit{fluctuations} in the shear stress should be maximized when frictional contact networks alternate the most between percolating and not-percolating. 

\section{Local Structure Evolution}

\begin{figure}
\includegraphics[width=\columnwidth]{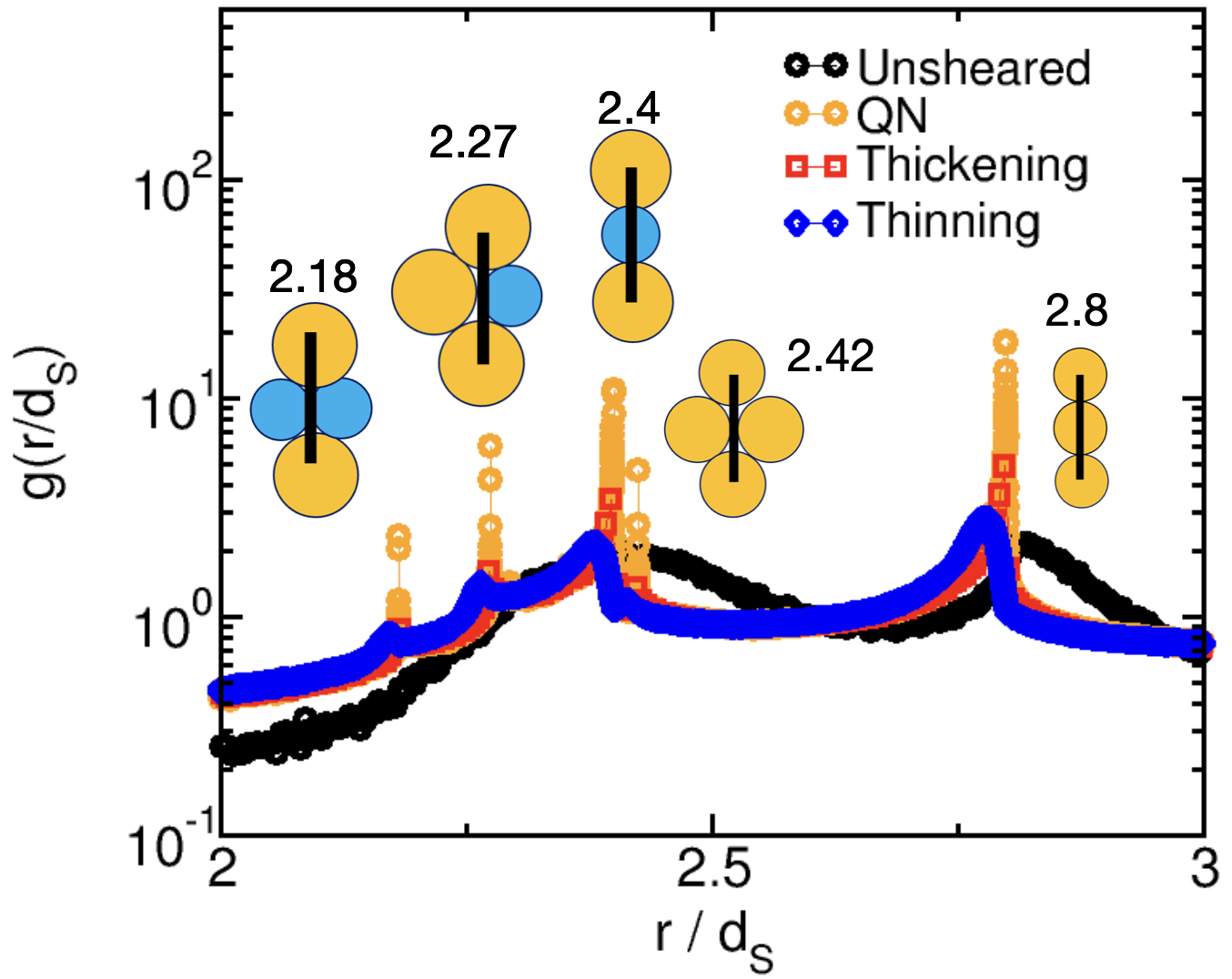}%
\caption[]{Evolution of g(r) from an un-sheared state (black) to {QN} (orange), thickening (red), and thinning (blue) flow. For an un-sheared system, g(r) shows broad peaks at distances corresponding to triplet chains. When shear is applied, sharp peaks corresponding to triplet chains and quadrilaterals emerge. Little change in the microstructure is seen between the {QN} and thickening flow while significant change occurs when the system begins to thin. {Shear rates are the same as in Figure \ref{fluctuations}.}}
\label{pcf_sheared_comp}

\end{figure}

\begin{figure}

\includegraphics[width=\columnwidth]{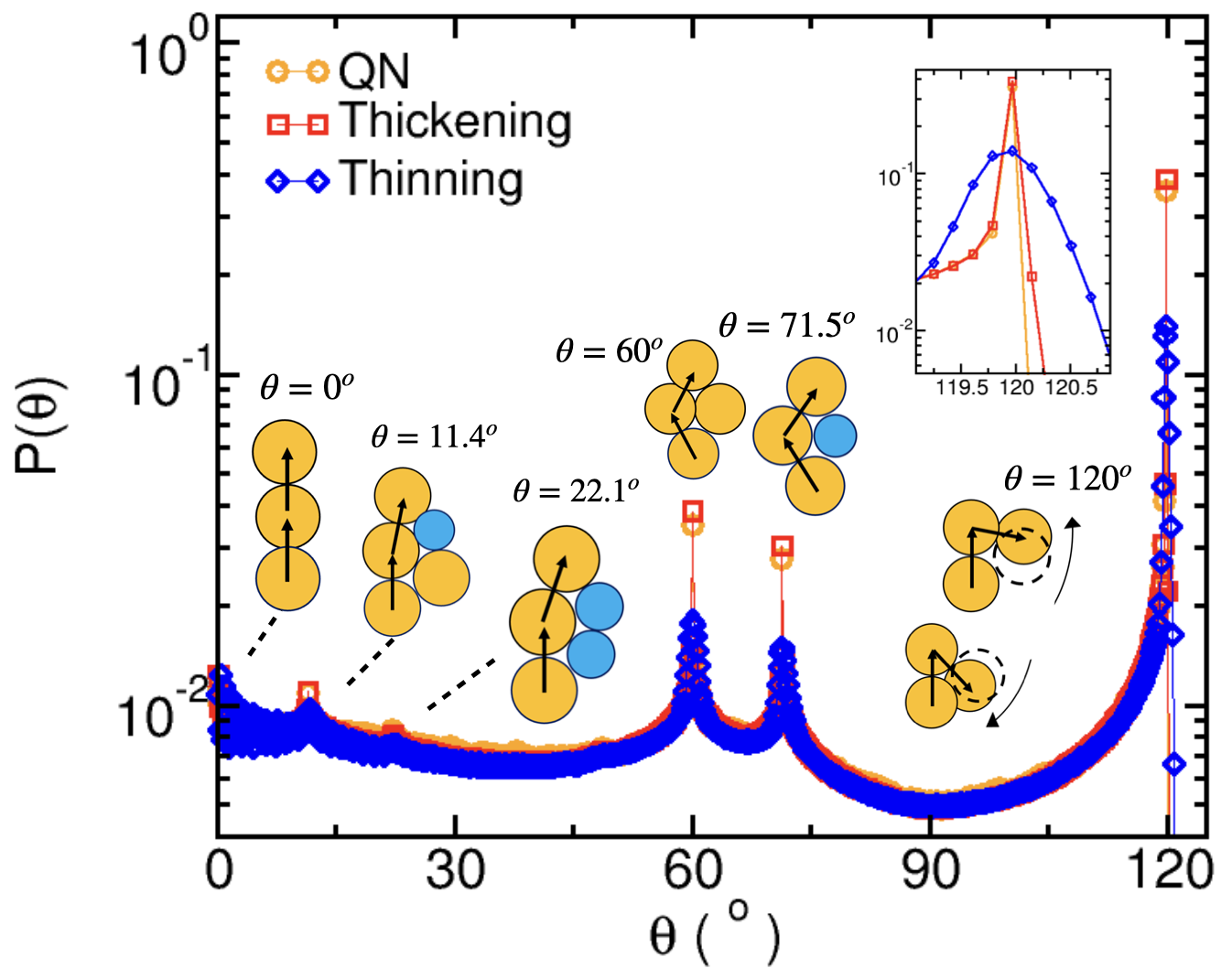}
\label{triplet_angle_dist}
\caption[]{The probability density distribution of $\theta$ at shear rates from the different flow regimes. Similar to the behavior of g(r), the $\theta$ peaks decrease and broaden significantly during the thinning regime. This change in the peaks is consist with chains {deforming} or triangles compressing, as shown by the enlarged picture of the $\theta = 120^o$ peak in the inset. {Shear rates are the same as in Figures \ref{fluctuations} and \ref{pcf_sheared_comp}.}}
\label{theta_def}
\end{figure}

We start by tracking how particle microstructure evolves with shear rate by calculating the pair correlation function $g(r)$, which describes the number density of particles in the system at a distance r from a reference particle, and is defined in Doi \cite{Doi_2013} as 

\begin{equation}
g(r) = \frac{1}{N}\sum_{m,n} \langle 
\delta (\vec{r} - \vec{r}_{mn}) \rangle     
\end{equation}

\noindent where $\vec{r}_{mn} = \vec{r}_m - \vec{r}_n$ is the relative position of particles $m$ and $n$ and $N$ is the total number of particles. One can compute $g(r)$ for each of the two particle species separately or together, and overall find similar features. Figure \ref{pcf_sheared_comp} shows $g(r)$ computed for the large diameter particles with an un-sheared state as a reference. When  steady flow has developed at a shear rate in the {QN} regime, we note instead sharp peaks in $g(r)$ that are not present in the un-sheared system. These peaks correspond specifically to local configurations of groups of three or four large particles or large and small particles as drawn in the figure. We note that these local structures are suggestive of building blocks of chain like structures, capable of growing to rapidly span the whole sample and efficient at transmitting stresses. We therefore hypothesize that they are building blocks of larger assemblies underpinning force chains, similar to what happens in jamming of dry grains under shear \cite{Bi_2011}, but here in steady flow, and therefore preferentially oriented in the compression direction. Interestingly, these structures emerge immediately under flow, i.e. even when the flow is {QN}, and during thickening there is little change in the $g(r)$ profile - we find only a slight decrease in the sharp peaks as the shear rate is increased (Figure \ref{pcf_sheared_comp}). However, the comparison in Figure \ref{pcf_sheared_comp} of the same $g(r)$ across the different flow regimes shows that with the onset of the thinning regime the peaks significantly decrease and broaden, suggesting that major changes develop in these structural building blocks. 

To understand how the local structures identified by $g(r)$ differ during thickening and thinning flows, we have quantified and tracked their changes. In particular, if large-diameter disks $i,j,k$ are connected such that $i$ touches $j$ and $j$ touches $k$, with $\vec{r}_{ij} = \vec{r}_{j} - \vec{r}_{i}$, we can characterize the shape of the $i,j,k$ triplet with the angle  

\begin{equation}
\theta = cos^{-1} ( \hat{r}_{ij} \cdot \hat{r}_{jk} )
\end{equation}

As shown in Figure \ref{theta_def}, $\theta = 0^o$ corresponds to a straight chain whereas $\theta = 180^o$ corresponds to an equilateral triangle. Tracking the probability density distribution of $\theta$, we find that, similar to $g(r)$, there is little change during thickening in the shape of the structural elements already emerged under shear, indicating that those structural elements, and their assembly, remain stable. However, the peaks of the distribution of $\theta$ evolve as thinning kicks in, indicating how the shape of these structures may be changing: the significant broadening of the peaks is consistent with disk triangles being compressed and with the {deformation} of initially straight triplets of large particles. If straight triplets are building blocks for force chains that carry stresses during thickening, the thinning must therefore take place through deformation and breakup of those stress bearing structures. 

The emerging picture is that structural elements formed under flow grow into percolating structures that can bear and transmit large stresses during thickening, as they allow for large contact forces to develop. The structural disorder and the pronounced heterogeneities in these percolating structures probably allow for the accumulation of large contact forces on a relatively small fraction of all the contacts, leading to the long-tail distributions shown in Figure \ref{fluctuations} (b) and hence (a). While we have evidence of distinctive local structures formed under flow and of their changes between shear thickening and thinning flow, quantities like $g(r)$ are not adequate to capture their organization into larger scale structures, potentially spanning and underpinning force chains. In spite of several attempts, identifying these larger structure by starting from the local structural elements and following their evolution remains very difficult and inconclusive, due to the complexity of the microstructural evolution even in this relatively simple two dimensional case. We have therefore focused on identifying the large scale assemblies that control the rheological changes signaled by the flow curve by directly using the information on the large contact forces, as explained in the following.


\section{Strong Force Network Percolation}

To further test the interpretation of our results, and identify the stress-bearing structures that control the rheology during thickening, we now combine the information on the tail of the contact force distributions contained in Figure\ref{fluctuations}(b) with the idea that specific subsets of the contact network may be more relevant to the rheology, as suggested for example in Goyal et al. \cite{Goyal_2024}. First, using the information contained in Figure \ref{fluctuations} (a) and (b), we have studied groups of disks whose contacts bear forces of magnitude $f$ which satisfies the condition

\begin{equation}
f \geq k f_{peak}
\label{threshold}
\end{equation}

\noindent where $f_{peak}$ is the peak of the contact force distribution for that shear rate. $f_{peak}$ represents the most probable force at a given shear rate, and the parameter $k$ can be used to adjust the force threshold in the condition above, with higher values allowing us to identify subsets of the contact network that include contact forces of higher magnitude. Here and in the following we call such a subset the {\it strong force network} (SFN), which is inspired by the approach proposed in Bi et al. \cite{Bi_2011} for dry granular materials under shear. These SFN's are networks of disks that are each over-constrained by large forces. In our analysis we have tested all results with several values for $k$, ranging from $0.75$-$1.25$, finding that they do not depend qualitatively on the value of $k$. We have then considered specifically the SFN made up of particles with at least 3 contacts above the threshold defined by Equation \ref{threshold}. Combining the information on constraint geometry (i.e. number of contacts) with the contact forces (making sure we focus on large contact forces that emerge during thickening) appears to be central to understanding the changes in the rheological behavior across the different flow regimes.

The choice to analyze 3 contact SFN's (3-SFN) is significant for a couple reasons. First, setting a minimum contact number of 3 ensures that the disks are locally over-constrained by strong forces (in the sense of Equation \ref{threshold}). Secondly, if we do not filter contacts by their force strength, we find that large networks of particles with 3 or fewer contacts always exist and percolate across the system. Meanwhile, networks of particles with 4 or more contacts never become large enough to percolate. 

\begin{figure}

\includegraphics[width=0.9\columnwidth]{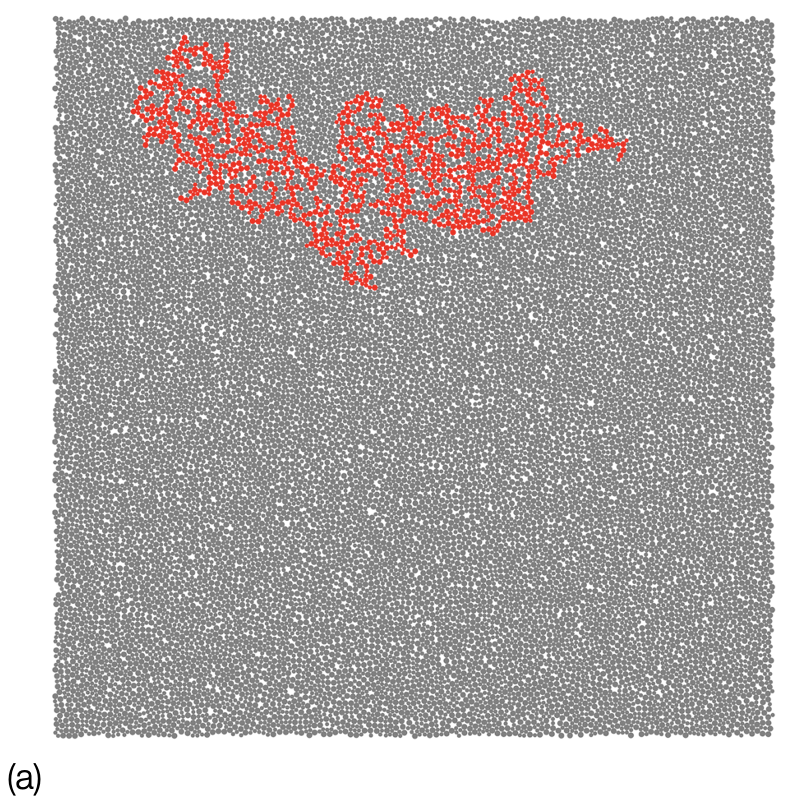}
\hfill
\includegraphics[width=0.9\columnwidth]{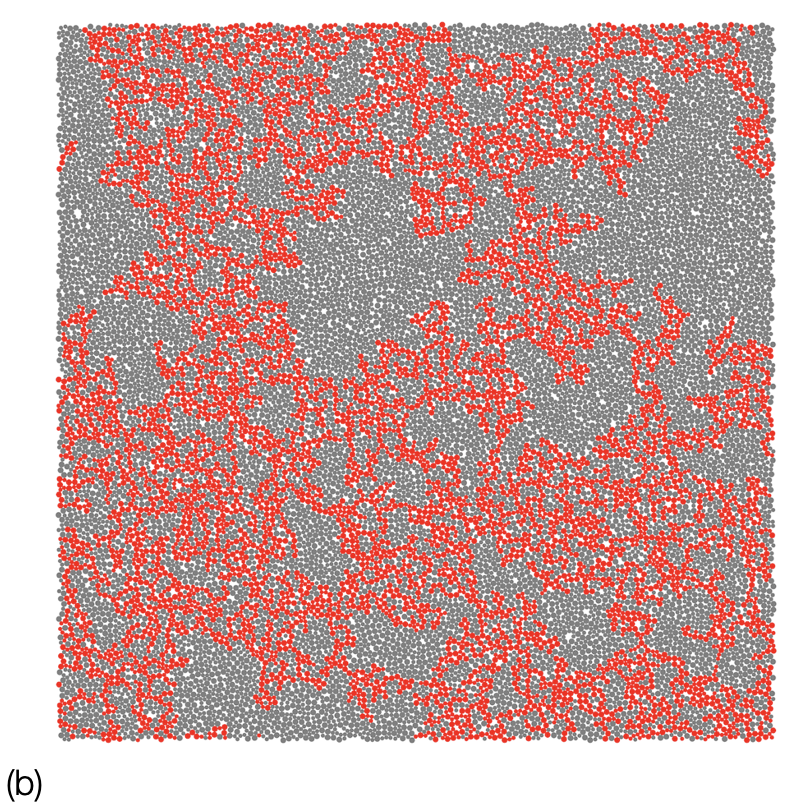}%

\caption{Filtering the networks of connected particles by the strength of the contact forces reveals heterogeneous percolation behavior. These strong force networks (SFN's) consist of particles with at least $m$ contacts with forces $f > f_{peak}$ where $f_{peak}$ is the second peak (first non-zero peak) of the force distribution. (a) and (b) show the largest cluster with $m=3$ at different times during the same experiment {for $\dot{\gamma}'= 5 \times 10^{-1}$}. All other clusters are hidden for clarity. Notably, the system shear stress in (a) is smaller than the shear stress in (b).}
\label{strong_force_networks}

\end{figure}

\begin{figure}
	\centering		
	\includegraphics[width=\columnwidth]{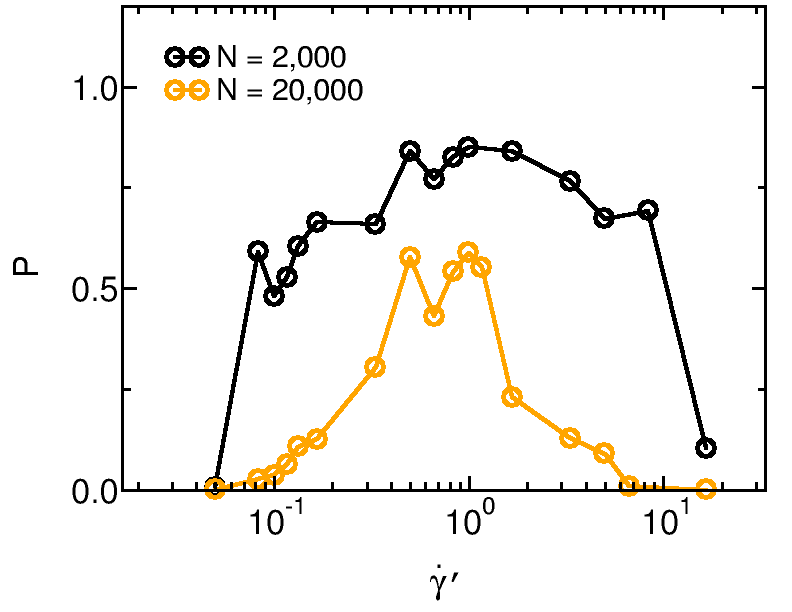}
	\caption{Strong force network percolation probability versus shear rate {for different system sizes, N}. The probability indicates the fraction of measured configurations in which a percolating network was found. For the system size of 20,000 particles, probabilities were calculated using 10,000 configurations evenly spaced in strain during which the system was subject to a strain of $\Delta \gamma = 10$.}
        \label{perc_probs}
\end{figure}

By analyzing the particle trajectories in steady flow, we can measure how the 3-SFN continuously percolate and break down under flow. Percolation here is defined as occurring when a specific 3-SFN spans both directions of the system. During thickening the 3-SFN experience dramatic fluctuations in their size and do not always percolate, as shown in the examples of Figure \ref{strong_force_networks}. To quantify how often the 3-SFN do percolate, we calculate a percolation probability - the fraction of configurations, over the whole time series, in which there is a percolating 3-SFN - and find that during thickening this probability strongly increases with the shear rate. As seen in Figure \ref{perc_probs}, the percolation probability grows with shear rate before peaking at the end of the thickening, to then decrease during thinning. We note that the coupling between the percolation of 3-SFN and the shear thickening/thinning becomes much less clear in small samples, where the data show instead a percolation probability very similar (and relatively large) across nearly the whole range of shear rates, without changing much between thinning and thickening. These data suggest that, if a percolation transition of 3-SFN underpins the shear thickening regime, only the large system size would provide enough resolution to accurately characterize it. While the full study of the percolation transition itself is beyond the scope of this paper, we have computed the cluster size distribution of 3-SFN from the time series in the thickening regime and found that the data for the large system size are consistent with proximity to a percolation transition, because of a power-law tail that clearly extends over more than one decade, with an exponent similar to the one predicted for random percolation in 2d \cite{Stauffer_2003} (see Supplementary Information, Figure 4). The data show the coexistence of several small and finite clusters with percolating ones during thickening, which is a hallmark of a percolation-type of growth.

\begin{figure}
\includegraphics[width=\columnwidth]{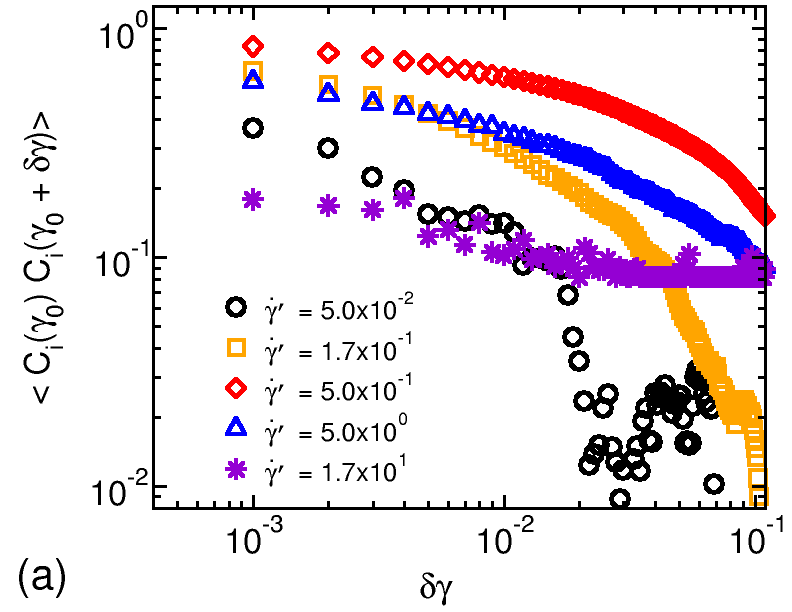}
\hfill
\includegraphics[width=\columnwidth]{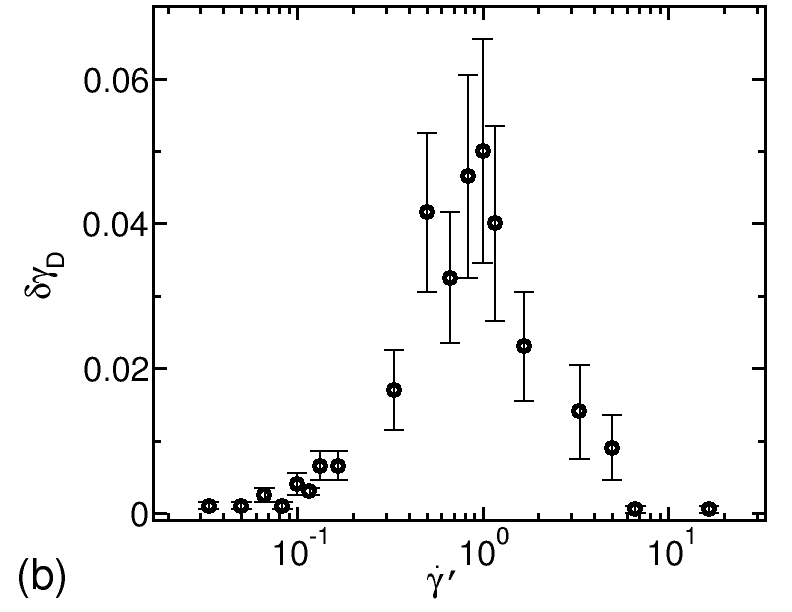}

\caption{As the 3-SFN's begin to percolate more frequently particles tend to stay in a percolating network longer. (a) The ensemble averaged autocorrelation of a particle's affiliation with a percolating network as a function of a change in strain $\delta \gamma$. (b) The de-correlation strain (the amount of strain it takes for the auto-correlation to fall below $\approx 1/e$) grows during the thickening regime and decreases during the thinning, consistent with percolation and de-percolation transitions. The error bars in the figure have been obtained considering the correlation to fall below $1/e \pm 0.05$.}
\label{percolation_transition}

\end{figure}

The data for the large systems in Figure \ref{perc_probs} clearly indicate that percolating 3-SFN are predominant during thickening and become rare in the thinning flow. The {deformation} of the local particle structures that we discussed in the previous section are likely responsible for interrupting, during thinning, the percolation of the structures that were instead able to efficiently transmit stresses during thickening, leading to a net increase of the viscosity of the suspension with increasing shear rate. 
The findings just described provide further insight when we measure the amount of strain through which a particle remains in a percolating 3-SFN, which also grows during thickening. At any given time or strain, any given disk can either belong or not to a percolating 3-SFN. This status is represented through a variable $C_i(\gamma_0)$, which corresponds to the affiliation of particle i at strain $\gamma$ to a percolating 3-SFN and is either 0 (belongs) or 1 (does not belong). We compute the time autocorrelation function

\begin{equation}
\langle C_i(\gamma_0) \; C_i(\gamma_0 + \delta \gamma)\rangle,
\end{equation}

\noindent which, averaged over all particles and strain windows of the same size at a given shear rate, is plotted in Figure \ref{percolation_transition}(a) as a function of strain. The results show how increasing the shear rate during thickening causes disks to stay in a percolating 3-SFN for longer strains. Conversely, increasing the rate during thinning decreases the strain window spent in a 3-SFN. The data therefore suggest that percolating 3-SFN become longer and longer lived during thickening and provide further evidence that they are a specific ingredient of the thickening flow. We can use these autocorrelation functions to estimate the characteristic strain $\delta \gamma_D$ over which a percolating 3-SFN persists on average, as the strain by which the average autocorrelation falls below $\approx 1/e$. The error bars in the plot refer to the autocorrelation falling below $1/e \pm 0.05$. $\delta \gamma_D$ is shown as a function of the shear rate in Figure \ref{percolation_transition}(b) to directly relate its rate dependence to the different regimes identified in the flow curve (Figure \ref{flowcurve}) and discussed so far.
The data clearly confirm that, during the thickening, not only do the 3-SFN's percolate more often, but also disks tend to stay in percolating networks over longer amounts of strain. During the thinning, instead, $\delta \gamma_D$ decreases with increasing the rate, confirming that the thinning flow, where the building units of stress-bearing structures become deformed, correspond to predominantly non-spanning 3-SFN. 
These results start to clarify that similar contact network structures could be found in both thickening and thinning, however their persistence to growing strains can be dramatically different due to the ability of their elementary building blocks to accumulate and transmit stresses, resisting the flow.  

We would like, at this point, to gain further insight into how the large scale assembly of the microstructural units induced by the flow may grow during thickening and control the fluctuations in the shear stress.

\begin{figure}
\includegraphics[width=\columnwidth]{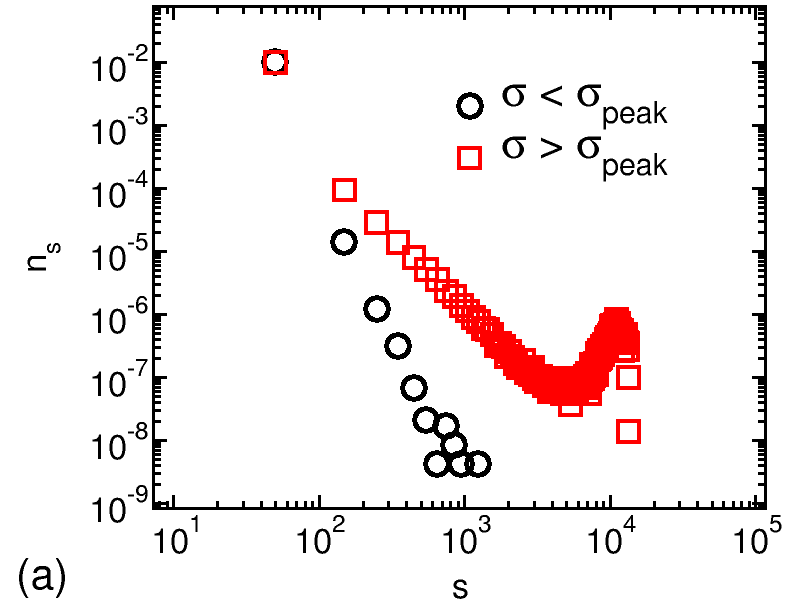}%
\hfill
\includegraphics[width=\columnwidth]{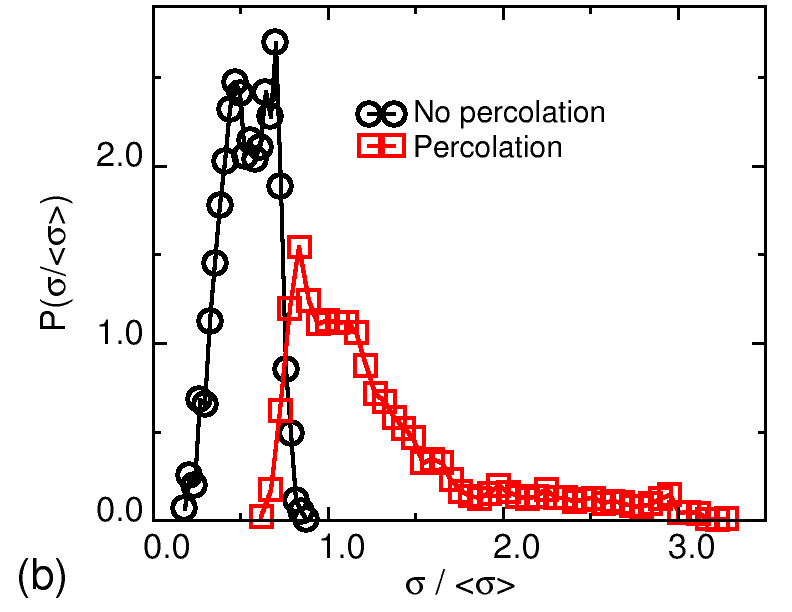}

\caption{(a) The cluster size distribution for the two stress states showing how larger clusters occur at higher stresses. The peak around $10^4$ for the $\sigma > \sigma_{peak}$ distribution corresponds to 3-SFN that encompass most of the particles and are system-spanning. (b) The stress distribution at times when the 3-SFN does and does not percolate. The stress values are rescaled by the ensemble average of the stress (calculated over all time). The stress is higher when the 3-SFN percolates. Data for both plots taken during the thickening ($\dot{\gamma}' = 5 \times 10^{-1}$).}
\label{CSD_stress_dists}

\end{figure}

\section{3-SFN Percolation and Shear Stress Fluctuations}
\label{strong-force}

The size of the 3-SFN bears significant correlations with the fluctuations of the shear stress in time, during the thickening regime. This correlation remains but grows notably weaker during the {QN} and thinning flow. Figure \ref{CSD_stress_dists}(a) shows the distribution of sizes of 3-SFN obtained from microscopic configurations for the steady state flow at a shear rate in the thickening regime ($\dot{\gamma}'=5 \times 10^{-1}$). The number $n_s$ of 3-SFN that have a number of disks $s$ is divided by the total number of particles in the system as outlined in Stauffer \cite{Stauffer_2003}. This is plotted against the size $s$, indicating that large scale assemblies of contacts that carry large forces are present and their maximum size grows with the rate. We then split the particle configurations, obtained for the same shear rate, based on whether the shear stress in the system is above or below the peak of the stress distribution $\sigma_{peak}$ in Figure \ref{fluctuations}(a). Figure \ref{CSD_stress_dists}(a) reveals that an increase in the frequency of the largest 3-SFN sizes clearly occurs in configurations where the shear stress exceeds the mean stress at that shear rate. {The peak around $10^4$ particles for the $\sigma > \sigma_{peak}$ distribution corresponds to 3-SFN that encompass most of the particles and are system-spanning. The difference in the shape of the two distribution seems compatible with large percolating clusters for $\sigma > \sigma_{peak}$ and non-percolating finite clusters for $\sigma < \sigma_{peak}$.} 

\begin{figure}

\includegraphics[scale=0.3]{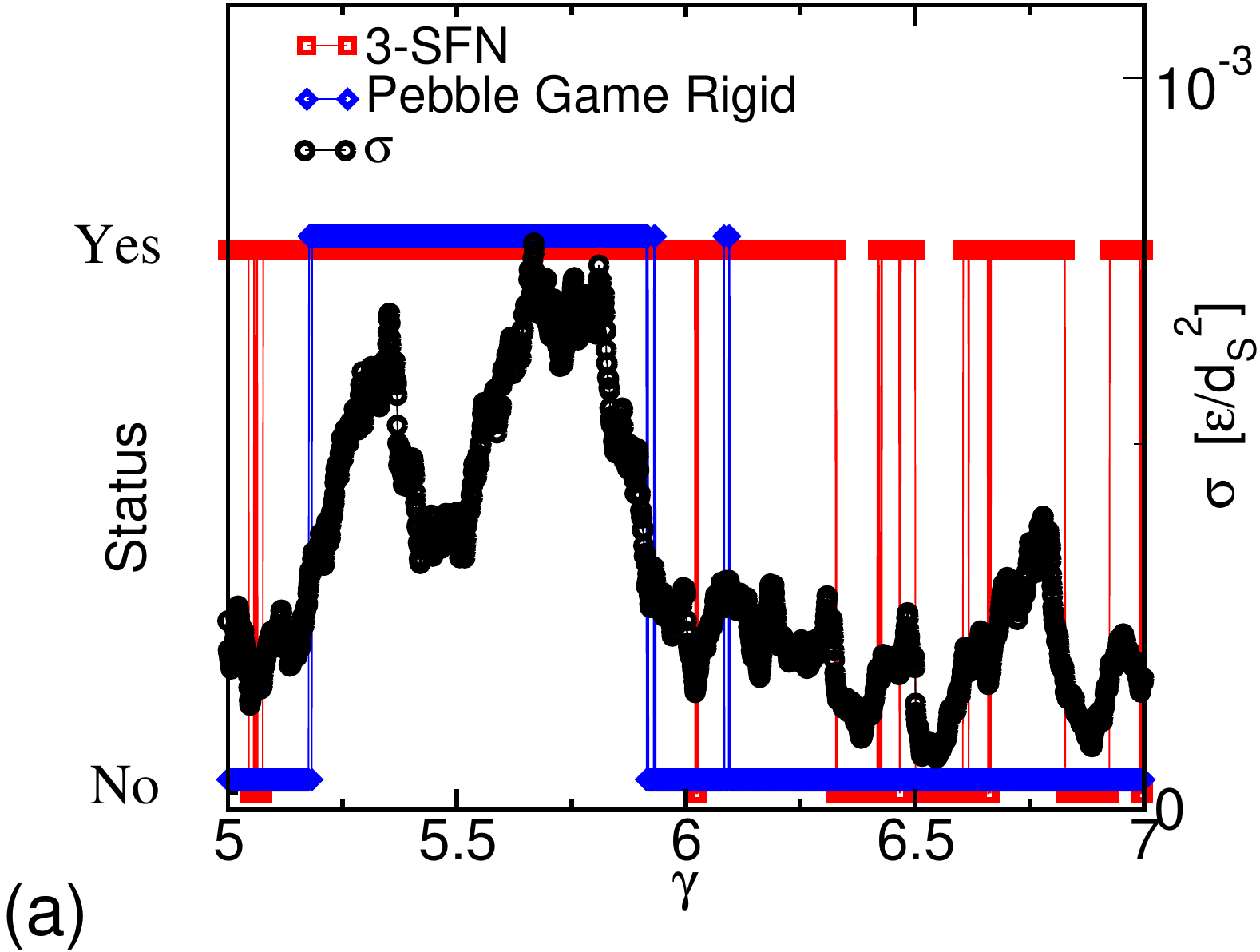}
\hfill
\includegraphics[scale=0.3]{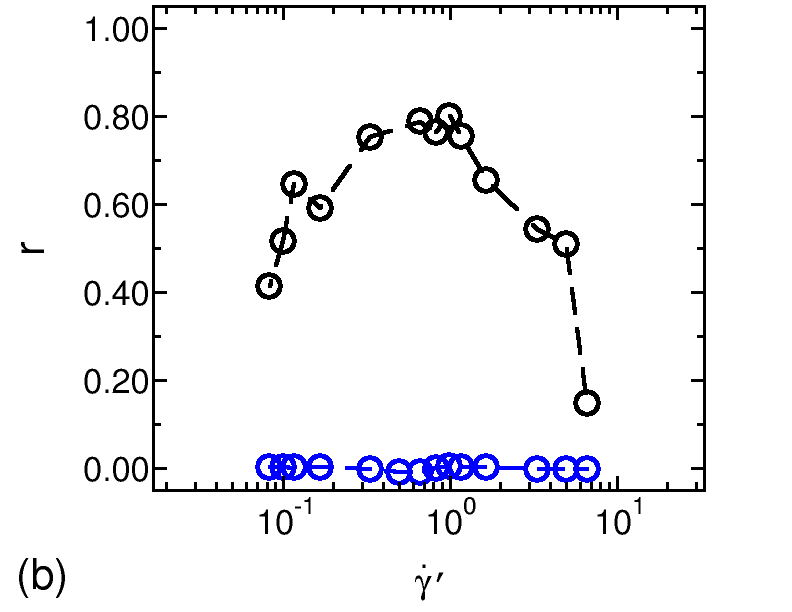}%

\caption{(a) Percolation status of the largest 3-SFN and pebble game cluster alongside the stress in the system. Shear rate is from the thickening regime ($\dot{\gamma}'=5 \times 10^{-1}$). For higher values of the stress the 3-SFN percolates while dips in the stress are associated with de-percolation. (b) The correlation between the percolation status of the 3-SFN and the stress versus shear rate {(black data)}. {These correlation values can be compared to the correlation between the stress with a random time series (blue data).} The thickening is associated with a sustained high correlation between 3-SFN percolation and shear stress while the {QN} and thinning flow are associated with increasing and decreasing correlation respectively.}
\label{perc_stress_plots}

\end{figure}

More evidence that the growth of 3-SFN is connected to the total shear stress is given in Figure \ref{CSD_stress_dists}(b). Here the data for the distribution of the total shear stress values in steady state of Figure \ref{fluctuations}(a) is split based on whether on not a percolating 3-SFN can be found in that configuration. We find that the presence of a percolating 3-SFN significantly shifts this distribution towards larger stresses. The results of Figure \ref{CSD_stress_dists}(a) and (b) suggest that the 3-SFN may be predictors of whether the shear stress in a given configuration is high or low compared to its most probable value ($\sigma_{peak}$). Figure \ref{perc_stress_plots}(a) explores this for a shear rate in the thickening regime. The status of whether the system at a given strain contains a percolating 3-SFN (or a percolating rigid network as discussed below) is plotted alongside the total shear stress. To quantify how predictive changes in the percolation of 3-SFN is of changes in the shear stress, we compute the Pearson correlation coefficient between the two:

\begin{equation}
\label{pearson_correlation}
r = \frac{\sum (\sigma(\gamma) - \bar{\sigma}) (S(\gamma) - \bar{S})}{\sqrt{\sum(\sigma(\gamma) - \bar{\sigma})^2 \sum(S(\gamma) - \bar{S})^2}}
\end{equation}

Here $\sigma(\gamma)$ is the total shear stress at strain $\gamma$ and $S(\gamma)$ is the percolation status of the 3-SFN being considered, with 0 indicating no percolation and 1 indicating percolation. The correlation coefficient is then plotted as a function of shear rate in Figure \ref{perc_stress_plots}(b) which shows that during the thickening regime there is reasonably good correlation (going up to $\approx 0.8$) between the percolation status of 3-SFN and the shear stress in the system. For reference, we also show in the plot the Pearson coefficient between the same time series of the shear stress and a randomly fluctuating variable.

 While the data provided here do not analyze all possible networks or possible subsets of the contact network, combined with the microstructural analysis discussed above they support the idea that 3-SFN and their percolation are distinctive ingredients of thickening in our model suspension. The 3-SFN identify the subset of the contact network that includes large contact forces developed upon thickening because their building blocks efficiently transmit stresses across the sample when they percolate. Finally, their percolation/de-percolation creates/interrupts the build-up of large contact forces and therefore underpins the large fluctuations of the total shear stress. Once the structural elements that compose them start to {deform}, efficient stress transmission is compromised and percolating 3-SFN can only survive over smaller and smaller strain windows. The {deformation} of the building blocks, as well as the more continuous percolation/de-percolation, help redistribute contact forces more homogeneously, reducing significantly the stress fluctuations over time and allowing the system to more efficiently dissipate energy, hence allowing for shear thinning. 
 
\section{Rigid Structures in the Contact Networks}

Our data suggest that the capacity of 3-SFN to transmit stress makes their percolation important for the system rheology. That is, the sensitivity of the percolation/de-percolation of 3-SFN to $\sigma$, as shown in Figure \ref{perc_stress_plots} (a), must be related to the rigidity of the percolating 3-SFN. At rest, and in two dimensions, rigid structures are usually effectively identified by the pebble game algorithm \cite{Jacobs_1997}, which accurately accounts for all constraints acting locally and globally on the microscopic degrees of freedom of a given particle configuration. Since the algorithm allows for different particle interactions, it has been effectively utilized also for colloidal gels \cite{Zhang_2019} and for sheared dry granular materials in static conditions \cite{Henkes_2016,Vinutha_2019,Babu_2023,Hor_Dashti}. 
Recent studies have shown that this approach can identify rigid structures relevant to shear thickening of dense suspensions \cite{van_der_Naald_2023, Santra_2024}, suggesting that it can capture well enough the structural self-organization underpinning stress transmission even under flow, in spite of not explicitly including moving boundaries and the effect of continuously increasing strain. Comparing therefore the percolating 3-SFN discussed so far with the rigid percolating networks detected by the pebble game algorithm could allow us to deeper understand the growth and optimization of stress-bearing structures during thickening.

Here we have applied a (3,3) pebble game algorithm to our model suspensions, to directly compare the rigid structures identified in this way, at specific shear rates corresponding to the different rheological regimes, with those emerging from our 3-SFN analysis in Figure \ref{perc_probs}. It should be noted that the pebble-game clusters are calculated by assuming that the system has already fully achieved rigidity, which may limit their capacity to capture the development or emergence of rigidity. We also note that the pebble-game algorithm considered here is the same used for shear jamming, i.e. it does not take into account the moving boundary conditions imposed by the steady-state flow. The networks identified by the pebble game algorithm appear to percolate at all rates studied here, independent from the specific rheological regime, therefore suggesting that they do not necessarily capture the structures responsible for the stress transmission. The comparison in Figure \ref{perc_stress_plots}(a), however, shows that during the thickening the percolation/de-percolation over time of both types of networks follows the fluctuations of the shear stress, while 3-SFN percolation may be better coupled to small stress fluctuations. 

A more direct analysis (Supplementary Information, Figures 5-7) further shows that there is a strong overlap between the two types of structures when percolating 3-SFN may occur but are not persistent (i.e., during the quasi-Newtonian and the thinning regime). However, when percolating 3-SFN become more persistent during the thickening regime, they are also heterogeneous and seem to contain a significant fraction (e.g. $\sim 50\%$) of particles that are not rigid according to the pebble game algorithm. Size seems to be a predictor of whether a percolating 3-SFN will also be rigid according to the pebble game during the thickening - 3-SFN containing around $60\%$ of all the particles are generally mostly rigid according to the pebble game algorithm and those with less particles are not. Comparatively, the percolation of rigid clusters identified by the pebble game algorithm shows a less consistent degree of correlation (i.e. slightly lower and less consistent Pearson coefficient) with the time series of the shear stress in the thickening regime. 

Overall it seems possible and consistent with the data provided here, that the 3-SFN may be better at capturing the rate dependence and the stress fluctuations because they are more likely to capture the structures responsible for the stress transmission. It would be useful in future work to perform further and extended analysis of similarities and differences of the two types of structures, sampling more rates and across different models.

\section{Conclusions}
\label{conclu}

Through numerical simulations of a minimal 2d model for dense suspensions under flow, we have identified specific subsets of the contact network that display significant changes depending on the different flow regimes. Flow helps the system self-organize into local structures that can rapidly grow and span the whole sample. Portions of the contact network that include disks with at least three frictional contacts (rigid or over-constrained) and where large contact forces develop tend to percolate during thickening. Their percolation/de-percolation produces large stress fluctuations that stem from the localization of large contact forces. Interestingly, these subsets of the contact network, which we call 3-SFN, have some overlap with rigid structures identified by the pebble game algorithm and identified by other groups in shear thickening 2d model suspensions but seem to be more sensitive to changes in shear rate and flow regimes, and better at capturing the emergence of rigidity during flow. 

The minimal model chosen here has many limitations but allowed us to effectively disentangle different concurrent effects, and demonstrate the generality of the shear thickening phenomenology, with potential implications for a broad range of complex materials. One interesting aspect, with this respect, is that, in spite of the model oversimplified description of the solvent, the results clearly demonstrate that contacts are hierarchically organized under flow, building networks that allow for localization of large contact forces during thickening. 

The softness of the particles in the model leads to a thinning regime at higher shear rate, through which the system avoids (or delays) shear jamming. Thanks to this feature of the model, we were able to identify the {deformation} of the local structures initially created by the flow, resulting in the de-percolation of the large scale assemblies that they form. The {deformation} of those structural units also allows for a more homogeneous redistribution of contact forces, which reduces significantly the large stress fluctuations that characterize thickening. These findings further reinforce the idea that shear thickening emerges from the growth of large scale, spanning assemblies of specific structural elements, which allow for transmitting stresses and localizing large contact forces.

When attempting a comparison with experiments on shear thickening suspensions, physical phenomena such as significant solvent migration and the associated concentration fluctuations \cite{Rathee_2022,Moghimi_2025} indicate complexities that certainly can not be captured by our minimal computational model. However, the rapid fluctuations of percolating structures observed here could be relevant to experimental observations of dramatic and rapid fluctuation in boundary stress and particle configurations \cite{Rathee_2017,Rathee_2022,Moghimi_2025}. The softness of the particles in the model, which allows for significant overlaps at high shear, may be relevant to specific experimental systems where particle deformability can promote the growth of concentration fluctuations during thickening. 

An interesting avenue for future work would be to explore further the percolation transition of 3-SFN during thickening. In particular, the connection to the rigidity percolation transitions discussed in the literature \cite{Henkes_2016,Goyal_2024} and to other properties of rigid clusters \cite{Santra_2024} should be explored.

\section*{Author contributions}
W.C.J.B. performed all simulations and analysis. V.H.A. provided some of the codes. All the authors 
designed the research, discussed the interpretation and wrote the article.


\section*{Conflicts of interest}
{There are no conflicts to declare.} 

\section*{Data availability}
Data used to generate plots for this article are available at a
Zenodo Open Repository (https://doi.org/10.5281/zenodo.18002353). 


\section*{Acknowledgements}
This work was supported by the National Science Foundation (NSF DMR-2226485 and NSF STC COMPASS – Center for Complex Particle Systems, Award No. 2243104). All the simulations discussed in this work were performed using High-Performance Computing capabilities available at Georgetown University through Google Cloud Platform.



\balance


\bibliography{sources} 
\bibliographystyle{rsc} 

\end{document}